# Disseminação de mensagens DTN com base em grupos de interesses

Eric V. das Neves[1], Ronaldo N. Martins[1], Celso B. Carvalho[2], Edjair Mota[1]

[1]Instituto de Computação – Universidade Federal do Amazonas (UFAM)
69080-900 – Manaus – AM – Brazil

[2]Programa de Pós-Graduação em Engenharia elétrica – Universidade Federal do Amazonas (UFAM)
69080-900 – Manaus – AM – Brazil

{ericimw, ronaldo.nascimentomartins}@gmail.com,
ccarvalho_@ufam.edu.br, edjair@icomp.ufam.edu.br

***Abstract.*** *Recent works explore social characteristics of nodes to improve message delivery rate in Delay Tolerant Networks (DTN). This work uses machine learning techniques to create node groups organized by common interests. Messages are sent to target groups, and from there to the final destination. Simulation results using The ONE simulator show that the larger the group size the higher the message delivery rate, that reaches 100% in some cases. The paper also presents results related to the groups of interest such as message delivery rat, delivery delay and an average number of hops to deliver messages. The overall results indicate that group-based routing is a promising research filed.*

***Resumo.*** *Trabalhos recentes exploram características sociais dos nós com o objetivo de melhorar as entregas de mensagens em redes tolerantes a atrasos e desconexões. Este trabalho explora o uso de técnicas de aprendizagem de máquina para criar grupos organizados por interesses comuns. As mensagens são enviadas com destino aos grupos e somente depois são direcionadas ao destino dentro do grupo. Resultados gerados no simulador The One mostraram que o envio de mensagens a grupos possui altas taxas de entregas, chegando em alguns casos a 100% quanto maior o tamanho do grupo. O artigo também apresenta medidas relacionadas a grupos, tais como taxa de entrega de mensagens, atraso de entrega e número médio de saltos. Os resultados indicam que o roteamento baseado em grupos é um campo promissor de pesquisa.*

## 1. Introdução

Redes Tolerantes ao Atraso e Desconexões (DTN - Delay Tolerant Networks) é uma arquitetura de redes de computadores que leva em consideração a falta de conectividade entre os nós da rede, ocasionada pela mobilidade e qualidade dos enlaces, dentre outros. As principais características de uma DTN são atrasos longos e/ou variáveis, e frequentes desconexões, o que pode ocasionar a não existência de um caminho fim-a-fim entre um nó fonte e um nó destino [Oliveira et. al., 2007.    Na ausência de uma conexão, um nó DTN armazena permanentemente a mensagem. Quando uma conexão é restabelecida, o nó tenta repassar a mensagem para um nó mais próximo. Assim, de nó em nó a mensagem é conduzida até o seu destino final. Uma possibilidade de encaminhamento da mensagem é entregá-la sempre que ocorre um contato (proximidade entre dois nós). O inconveniente é que essa decisão consome recursos dos dispositivos móveis, tais como bateria e espaço em disco. Faz-se necessária a busca por repasse mais eficiente

Recentemente, algumas pesquisas procuram explorar a possibilidade de entrega de mensagens a partir da perspectiva das características sociais dos usuários [Gal et. al., 2009]. Nós acreditamos que a taxa entrega das mensagens pode crescer ainda mais se detectarmos grupos sociais. Uma vez detectado o grupo social a que pertence o usuário destino, a entrega de uma mensagem a um desses usuários potencializa a entrega da mensagem ao seu destino, pois é bem provável que usuários desse grupo se encontrem com mais frequência. A grande questão é a detecção ou descoberta de grupos. Técnicas de Aprendizagem de Máquina podem ser utilizadas para esse propósito.

Este trabalho investiga e propõe um mecanismo de descoberta de grupos de usuários em redes DTN e o agrupamento destes utilizando características sociais previamente estabelecidas. Utilizando dados de movimentação reais, implementamos o mecanismo utilizando a ferramenta The ONE Simulator [Keränen e Kärkkäinen, 2009], amplamente utilizada pela comunidade científica de redes DTN. Projetamos um conjunto de experimentos criteriosamente selecionados, para estimar a taxa de entrega de entrega de pacotes quando se direciona as mensagens a grupos de usuários.

Para tratar os assuntos comentados, dividimos o artigo nas seguintes seções: Na Seção 2 discutimos alguns trabalhos relacionados. A Seção 3 e 4 aborda os conceitos usados para entrega multicast e identificação de grupos de interesse. Na Seção 5, detalhamos a proposta do protocolo abordado neste trabalho. Na Seção 6, apresentamos a metodologia usada para a execução dos experimentos. Na Seção 7, apresentamos e discutimos os resultados experimentais. Finalmente, na Seção 8 apresentamos as conclusões e sugestões para trabalhos futuros.

## 2. Trabalhos relacionados

Os protocolos de roteamento DTN, em sua maioria, exploram a mobilidade dos nós, sejam eles pessoas ou coisas. Zhu et. Al. (2013) classifica e descreve em três categorias os esquemas de roteamento DTN, baseado em (i) um elemento da rede que leva e traz mensagens periodicamente (*message-ferry-based)*; (ii) um contato que ocorre oportunisticamente (*opportunity-based)*; (iii) previsões de contatos (*prediction-based)*. Uma abordagem mais recente (*social-based forwarding*), explora os aspectos sociais dos nós que, segundo Gao et. al. (2009), apresentam melhor desempenho quando comparado às abordagens tradicionais baseadas em heurísticas ou predições de mobilidade. Hui et. Al. (2006) estudaram a estrutura social da mobilidade humana propondo o protocolo BUBBLE Rap. Outros trabalhos foram desenvolvidos nesta linha, utilizando-se de outras características sociais como interesse do usuário, similaridade social e estrutura de comunidade. Surgiu, então, o termo *Socially Aware Networking* (SAN). Neste mesmo sentido, Xia et. al. (2016) propuseram o algoritmo *Int-Three* que explora o uso de características de interesse.

Grande parte dos trabalhos utiliza o envio de mensagens do tipo ponto-a-ponto para somente um destino final (*single destination*). Gao et. al., (2009) apresentaram um protocolo operando com múltiplos destinos (*multicast*) e utilizando a perspectiva social, porém sem o uso de técnicas de aprendizagem de máquina. Outro campo de estudo surgiu chamado *Social Network Analysis* (SNA), usado para a tomada de decisão no repasse das mensagens. Gao mostra duas observações importantes: (i) as comunidades, formadas naturalmente através das relações sociais, (ii) as centralidades que mostram haver nós de conhecimento comum a outros.

Paralelamente, o uso de técnicas de aprendizagem de máquina são usados em conjunto com SNA. Com esta ideia, Hajiaghajani e Biswas (2017) apresentaram o algoritmo *Q-Learning based Gain-aware Routing* (QGR), que utiliza aspectos sociais baseados na venda de cupons em lugares públicos usando técnica de aprendizagem por reforço.

A utilização de envio de mensagens DTN para um grupo de nós, na forma multicast, e a utilização de um algoritmo de clusterização, para a formação de grupos de interesses, ainda não foi explorada, sendo o objetivo deste trabalho.

## 3. Entregas Multicast para Grupos Sociais

Este trabalho utiliza técnicas de aprendizagem de máquina para criar grupos (*clusters*), associando a cada grupo um número definido de interesses sociais. Algoritmos de clusterização baseados em aprendizado de máquina, organizam dados em clusters. O algoritmo de clusterização Esperança-Maximização [Moon, 1996] pode encontrar o valor ótimo do número de clusters. O algoritmo K-Means organiza os dados em K clusters, sendo K escolhido. Como desejamos agrupar os nós em um grupo K (número de interesses) grupos, por esse motivo utilizou-se o algoritmo K-Means [Hartigan e Wong, 1979].

Em conjunto, utilizou-se o envio de mensagens com uma abordagem multicast. Dados reais obtidos de traces do evento Infocom [Scott et. al., 2006] e da universidade do MIT [Eagle e Pentland, 2005], foram utilizados no simulador The ONE para disponibilizar a movimentação dos nós nas simulações. Foram observados os comportamentos de variáveis como taxa de entrega de mensagens aos grupos e número de saltos percorridos pelas mensagens até alcançar os grupos de destino. A Figura 1 ilustra o principal objetivo da abordagem, nela dois nós de grupos de interesses diferentes desejam trocar mensagens, a linha laranja mostra como seria o roteamento unicast ideal. A linha verde mostra o objetivo deste trabalho, fazer com que a mensagem tenha como primeira meta alcançar o grupo de interesse (na Figura 1 o grupo do interesse 1 está contido na região delimitada pela linha verde) e em um segundo momento percorrer o grupo em busca do nó destino.

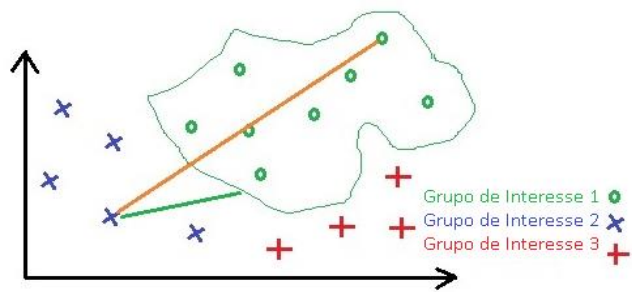

**Figura 1. Formação de grupos de interesse.**

## 4. Identificação dos Grupos de Interesse

É necessário identificar os nós pertencentes a uma lista conhecida de grupos de interesse. Diz-se que elementos do mesmo grupo ou cluster tem similaridade entre eles e dissimilaridade a elementos de outros grupos. Existem diversos algoritmos de clusterização na área de aprendizagem de máquina, dentre eles podemos citar k-means (MacQueen,1967), k-medoids (Kaufman & Rousseeuw, 1987). Nesta pesquisa utilizamos o algoritmo K-means para criar uma quantidade conhecido de clusters ou grupos. Assim, assumimos que conhecemos de antemão os número K de clusters.

## 4.1. Criação de Grupos – Clusterização usando K-Means

O algoritmo k-means constrói k partições dos dados e busca uma partição com k *clusters* que otimize o critério de particionamento escolhido. Possui uma função objetivo (1) que busca minimizar a soma dos quadrados das distâncias tal que:

$$E = \sum_{i=1}^{k} \sum_{p \in C_i} (p - m_i)^2 \qquad (1)$$

Onde:

$E$ é a soma dos quadrados dos erros para todos os objetos no dataset.

$p$ é o ponto no espaço representando um dado objeto;

$m_i$ é o centróide do cluster $C_i$.

### 4.1.2 Modelagem dos atributos

Neste problema, temos um conjunto de $i$ nós $N = \{N_1, N_2, ..., N_i\}$. Para cada nó $N_i$, pode-se associar um ou mais interesses. Tendo um conjunto de interesses finitos $I = \{I_1, I_2, ..., I_n\}$, cada interesse pode assumir um valor binário, onde, 0 significa não possui interesse ou 1 tem significado possui interesse. Então, podemos representar que o nó ($N_1$) possui k interesses, como $N_1<I_1, I_2, ...,I_k>$. Assim, $N_1<0,1>$, representa que o nó 1 esta associado a dois interesses, i1 e i2, mas só tem interesse em i2.

### 4.1.3 Algoritmo K-means

O algoritmo K-Means foi primeiramente descrito por Hartigan (1975), este trabalho utiliza a versão eficiente do algoritmo proposta por Hartigan e Wong (1979). Um dos objetivos do do k-means é encontrar similaridades entre os dados da amostra. O K, de K-Means, é a quantidade de centróides (pontos centrais dos grupos) que serão criados e auxiliará no objetivo de encontrar a similaridade dos dados. Inicialmente pode se escolher K pontos (centróides) aleatórios iniciais. A cada iteração novos centróides são calculados, estes deverão estar localizados na distância média entre os pontos do centróide. O algoritmo pára quando nenhum ponto precisa mudar de centróide associado. A seguir é apresentado o algoritmo K-means, onde D é um conjunto de pontos (cada ponto é um nó DTN).

---

**Algoritmo 1. Algoritmo de Clusterização**:

1: **procedimento** K-Means ($k, D$)
2: Passo 1: Selecionar arbitrariamente $k$ pontos do conjunto D, como os clusters iniciais
3: Passo 2: Calcular os centróides dos $k$ clusters da posição atual. //iteração corrente
4: Passo 3: Associar cada ponto ao centróide do cluster mais perto (maior similaridade)
5: Passo 4: Retornar ao passo 2 e parar quando não houver mais mudanças significativas entre os pontos.
6: **retornar** K centróides e os pontos de cada cluster.

---

Onde:

$k$ é o número de centróides.

$D$ é o conjunto de dados.

## 5. O Protocolo

O protocolo proposto neste trabalho se baseia em 3 processos para o repasse das mensagens até o grupo de destino: Classificação da mensagem gerada, agrupamento dos nós com interesse pela mensagem e repasse da mensagem entre os nós até alcançar o grupo e posteriormente alcançar o nó de destino dentro do grupo.

### A. Classificação da Mensagem

Cada mensagem gerada pelos nós de origem é classificada conforme as 35 categorias de interesses presente no trace Infocom.

$$M_n = I[k] \qquad (2)$$

Onde:

$M_n$ = Mensagem de índice $n$;
$I$ = Categorias de interesses;
$k$ = Índice de classificação referente ao interesse por área para classificação da mensagem.

Ex: dado $I = \{Content\ distribution, Power\ control, Service\ overlays\}$
$M_1 = I[3]$, onde a classificação da categoria de interesse da mensagem $M_1$ é a Service overlays.

### B. Agrupamento do Nós

Após a classificação da mensagem o índice de classificação da mesma é enviado para o clusterizador que agrupa os nós presentes na rede de acordo com seus interesses pela categoria da mensagem gerada e retorna esse conjunto de nós.

$$C = N | N_n\ possui\ interesse\ igual\ a\ I[k] \qquad (3)$$

Onde:

$C$ = Cluster de nós com o mesmo interesse;

$N$ = Conjunto de nós pertencentes a rede;

$n$ = Índice do nó pertencente ao conjunto da rede;

$I$ = Conjunto de interesses;

$k$ = Índice de classificação.

Ex: $C = [5,8,15,23,27,31]$

O retorno da resposta do clusterizador é o conjunto de identificadores (IDs) dos nós que possuem o interesse pela categoria de classificação da mensagem.

### C. Roteamento entre os nós

O repasse das mensagens do nó de origem ao grupo de destino acontece entre os nós que fazem parte do grupo escolhido. Isso acontece pelo fato de que os nós que foram agrupados possuem interesse pela mensagem, o que os elege como bons repassadores já que os nós adjacentes a eles também possuem interesse pela mesma mensagem.

**Algoritmo2. Interest Cluster Transfer**
1: InterestClusterTransfer(Cluster, NoA, NoB)
**2.**     **if**(NoB == Cluster[n]) **then**
2:        NoB.buffer = NoA.mensagem
3:     **else**
4:        encerraConexao(NoA, NoB)
5:     **end if**
6: **end procedure**

O Algoritmo 2, mostrado acima, recebe o conjunto de nós que foram agrupados (Cluster), o nó que possui a mensagem (NoA) e o nó para o qual se pretende repassar a mensagem (NoB). Ocorre uma verificação se o nó B está presente no conjunto de nós que foram agrupados, caso esteja presente, a mensagem é repassada a ele, se isso não for verdade, a conexão entre os nós é encerrada e a mensagem continua com o nó A.

## 6. Metodologia

Para avaliação do protocolo proposto neste trabalho, foram realizadas simulações usando o simulador The ONE Simulator. As simulações tiveram como base para a mobilidade dos nós, o uso dos arquivos de trace chamados respectivamente Infocom [ Pentland et. al 2009], e Reality [Chaintreau et. al 2006].

O trace de mobilidade chamado Infocom possui 337.418 segundos de dados coletados referente ao experimento realizado durante o evento Infocom 2006 em Barcelona, que durou 4 dias.

Este trace possui dados de contatos entre 4.725 nós juntamente com a classificação de assuntos de interesses, 35 categorias, que cada nó entre os IDs 21 a 99 se propôs a preencher em um formulário disponibilizando suas preferências por determinada categoria de assunto.

O trace Reality possui 16.979.816 segundos de dados coletados na universidade de Helsinki. Este trace possui 96 nós e seus respectivos históricos de contatos captados através de conexões entre os dispositivos via bluetooth.

Devido a limitada capacidade de armazenamento dos nós simulados, foi escolhido como política de gerenciamento de buffer o Drop Oldest, que descarta a mensagem mais antiga armazenada caso aconteça uma sobrecarga no recurso de armazenamento. Para implementação do algoritmo de clusterização foi usado no The ONE a biblioteca "weka.jar" juntamente com o algoritmo de clusterização K-means. O mesmo utiliza o conceito de aprendizado não supervisionada para fazer os agrupamentos, através da análise nos padrões nos dados de entrada tendo como parâmetro a distribuição no espaço.

O desempenho do protocolo proposto foi mensurado através das métricas de avaliação usadas amplamente em outros trabalhos, isto é:
- Taxa de entregas: a proporção de mensagens recebidas pelos nós de destinos em relação ao número de mensagens geradas nos nós de origem.

- Delay de entrega: o intervalo de tempo médio entre o evento de envio e o recebimento da mensagem.
- Número de saltos: número de nós intermediários que receberam e repassaram a mensagem até seu destino.
- Custo médio: o número médio de repasses por mensagens entregue ao destino.

Para a formação dos clusters pelo algoritmo K-means foi usado o fator "número de interesses por nós". Na avaliação da formação dos clusters foi usado as seguintes métricas:
- Número de nós no cluster: referente ao número de nós que formam o cluster gerado pelo K-means.
- Recurso usado: a relação entre os nós usados na formação do cluster e o total de nós da rede.

## 7. Resultados

Nesta seção apresentamos e discutimos os resultados obtidos nas simulaões realizadas.

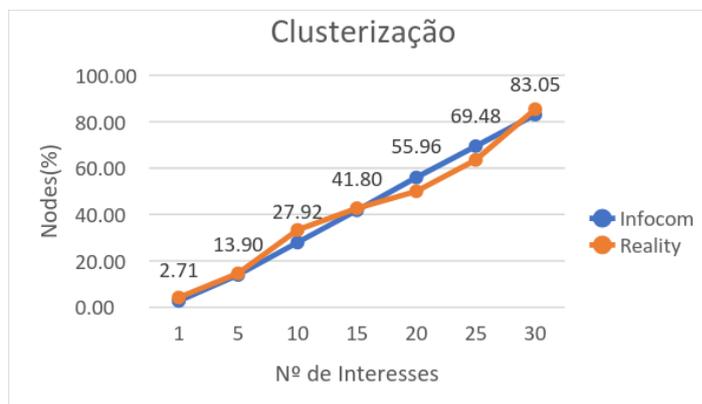

Figura 1. Nodes usados na clusterização por número de categorias de interesses.

D. Clusterização

A Figura 2 apresenta o resultado do processo de clusterização quando a quantidade de categorias de interesse existentes na rede varia de 1 a 30.

É possível observar que com o aumento do número de categorias, o clusterizador aumenta a porcentagem de nós da rede que são utilizados para compor os clusters, sendo que o número de nós é fixo. Com isto, mais nós são utilizados para o repasse das mensagens e consequentemente mais recursos da rede são utilizados. Esse comportamento está presente nos dois cenários analisados.

A formação dos grupos de nós exige da rede um aumento do uso de seus recursos proporcionalmente ao número de nós que compõem cada grupo. Se um grupo possui um número menor de componentes em relação ao número de nós total da rede, os recursos usados serão menores, assim como se o grupo formado for grande em comparação aos nós disponíveis, o uso de seus recursos também será usado de forma maior em relação a grupos menores.

E. Taxa de Entrega

Podemos verificar na Figura 3 a taxa de entrega realizada nos dois cenários utilizados em nosso trabalho. Assumimos que a entrega para determinado grupo de nós acontece quando um dos componentes do grupo recebe a mensagem destinada ao grupo. Nesse sentido nosso protocolo manteve uma alta taxa de entregas ao grupo de destino nos dois cenários.

Ao analisarmos a taxa de entrega pelo percentual de nós do grupo, podemos verificar que no cenário Reality a taxa ficou acima de 60% e abaixo de 75% de entrega para 1 e 5 categorias de interesses, enquanto que a partir de 10 categorias, a taxa de entrega se estabeleceu acima de 90%, e em alguns casos se aproximando de 100%.

No cenário Infocom, a taxa de entrega para 1, 5 e 10 categorias de interesses ficou abaixo de 20%, a partir de 15 categorias a taxa teve um aumento, ficando acima de 25% e mantendo-se. Na comparação dos cenários, o Reality manteve sua taxa de entrega superior ao Infocom. Isso se deve diretamente ao fato de que há uma maior interação entre os nós compreendidos entre os ids 1 a 99, o que equivale a 2% dos nós disponíveis no Infocom, o que não ocorre no cenário Cambridge, por possuir os contatos mais distribuídos em seu arquivo de trace.

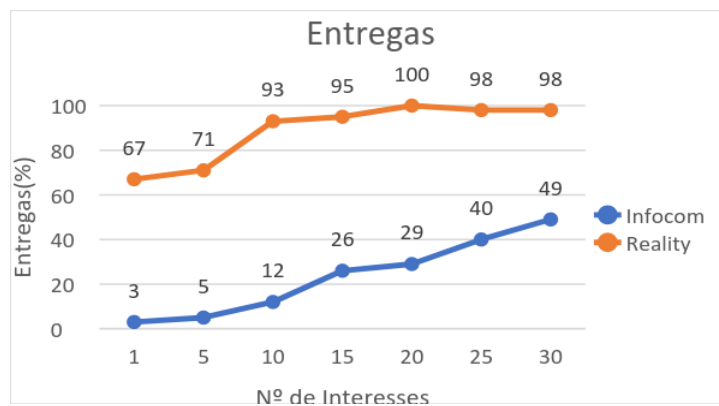

Figura 2. Taxa de entrega de mensagens por número de categorias de interesse.

Mesmo diante desse cenário de inconstância na interação entre os nós, nosso protocolo se manteve constante na taxa de entrega para os grupos e mostrou-se crescente no número de entregas conforme aumentava o número de categorias de interesse existentes na rede, categorias estas aos quais os nós poderiam se associar.

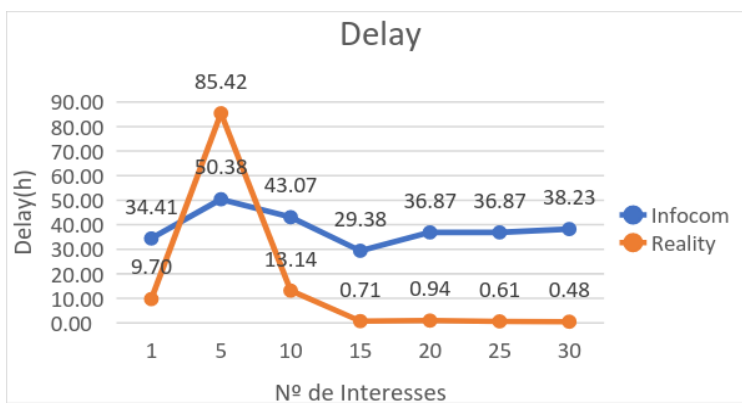

Figura 3. Delay por número de categorias de interesse

F. Delay Médio

É possível verificar através da Figura 4 que, conforme o número de categorias de interesse aumenta, o processo de clusterização faz com que mais nós compunham os grupos. Por esse motivo, a interação entre os nós que formam os grupos é maior, levando a diminuição do tempo de delay para entrega das mensagens ao grupo e aos seus componentes.

G. Custo Médio

A medida de custo médio é feita avaliando a relação de encaminhamentos feitos pelos nós da rede em relação ao número de mensagens geradas destinadas aos nós que compõem o grupo.

Podemos verificar na Figura 5 que conforme o número de categorias de interesses aumenta, ocasiona a formação de grupos maiores e, o custo médio de entrega de mensagens aumenta proporcionalmente. O custo médio do cenário Infocom se manteve superior ao cenário Reality devido ao número de nós que compõem seus arquivos de traces e consequentemente o número de nós que compõem seus grupos. O custo de recursos que deve ser consumido para repasse de mais cópias de mensagens a outros nós dentro dos grupos é proporcional ao tamanho do grupo formado pelo clusterizador.

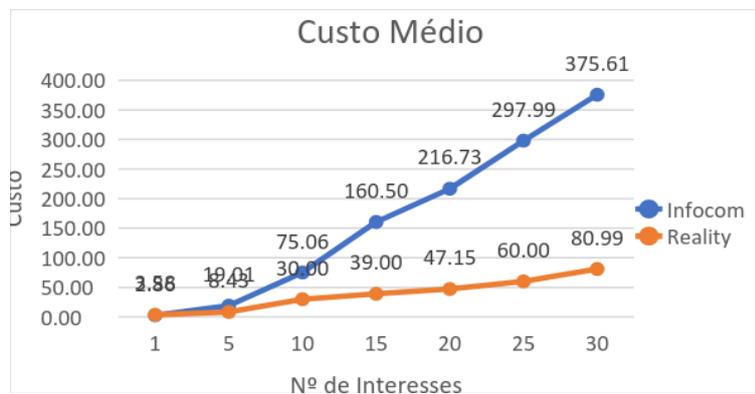

Figura 4. Custo médio por número de categorias de interesse.

8. Conclusões e Trabalhos Futuros

Neste trabalho propusemos um protocolo de disseminação de mensagens DTN baseado em grupos de interesses. Este protocolo considera o interesse dos nós em determinado tópico de assunto para agrupamento dos mesmos e assim tornar mais eficiente a entrega de mensagens a grupos em redes DTN e aos seus componentes.

O protocolo executa duas tarefas principais: a primeira tarefa usa a técnica de aprendizado de máquina para, clusterizar, formar grupos de nós com o mesmo interesse. A segunda tarefa se resume em repassar as mensagens aos nós que foram agrupados até que chegue aos nós destino.

Para determinar a eficácia de entrega das mensagens, usamos dois cenários para simular nossos experimentos, levando em conta de que cada cenário possui mais nós em relação ao outro.

Podemos perceber que os fatores de clusterização exercem uma grande importância para a formação dos grupos dos nós, o que por consequência tem bastante influência na taxa de entrega das mensagens aos grupos.

Concluímos que o protocolo proposto foi eficiente em relação a entrega a grupos de nós, sempre mantendo a taxa de entrega a pelo menos um nó pertencente ao grupo gerado. Em um cenário que possuía mais interação entre os nós componentes da rede, o protocolo se manteve sempre acima de 60% de taxa de entregas. Mesmo em cenários que possuíam menos interações entre seus nós, o protocolo foi capaz de entregar a mensagem ao grupo de destino, o que leva a conclusão de que com o acréscimo de tempo a mensagem chegaria aos seus destinos individuais.

Para trabalhos futuros, pretendemos usar múltiplos fatores de interesses, acrescidos de pesos de importância entre eles, para a formação dos grupos e melhorar a eficiência do roteamento intra e extra grupos através de adaptabilidade ao contexto social.

## 9. Referências

Chaintreau, A., P. Hui, et al. Impact of human mobility on the design of opportunistic forwarding algorithms. In Proc. INFOCOM, April 2006.

Daly, E., and Mads H. "Social network analysis for routing in disconnected delay-tolerant manets." Proceedings of the 8th ACM international symposium on Mobile ad hoc networking and computing. ACM, 2007.

Gao, Wei, et al. "Multicasting in delay tolerant networks: a social network perspective." Proceedings of the tenth ACM international symposium on Mobile ad hoc networking and computing. ACM, 2009.

Hartigan, J. A., & Wong, M. A. (1979). Algorithm AS 136: A k-means clustering algorithm. *Journal of the Royal Statistical Society. Series C (Applied Statistics)*, *28*(1), 100-108.

Hui, Pan, Jon Crowcroft, and Eiko Yoneki. "Bubble rap: Social-based forwarding in delay-tolerant networks." IEEE Transactions on Mobile Computing 10.11 (2011): 1576-1589.

Lindgren, A., Avri D., e Olov S. "Probabilistic routing in intermittently connected networks." ACM SIGMOBILE mobile computing and communications review 7.3 (2003): 19-20.

Keränen, A., J. Ott, and T. Kärkkäinen, "The ONE Simulator for DTN Protocol Evaluation," in SIMUTools '09: Proceedings of the 2nd International Conference on Simulation Tools and Techniques. New York, NY, USA: ICST, 2009.

Moon, T. K. The expectation-maximization algorithm. *IEEE Signal processing magazine*, *13*(6), 47-60. 1996.

Oliveira, C. T., Moreira, M. D., Rubinstein, M. G., Costa, L. H. M., & Duarte, O. C. M. Redes tolerantes a atrasos e desconexões. *SBRC Simpósio Brasileiro de Redes de Computadores e Sistemas Distribuídos*. 2007.

Pentland, A., N. Eagle, and D. Lazer, "Inferring social network structure using mobile phone data," Proceedings of the National Academy of Sciences (PNAS), vol. 106, no. 36, pp. 15 274–15 278, 2009.

Xia, Feng, et al. "Data dissemination using interest-tree in socially aware networking." Computer Networks 91 (2015): 495-507.

Xia, F., Liu, L., Li, J., Ma, J., & Vasilakos, A. V. "Socially aware networking: A survey." IEEE Systems Journal 9.3 (2015b): 904-921.

Zhu, Y., Xu, B., Shi, X., & Wang, Y. A survey of social-based routing in delay tolerant networks: Positive and negative social effects. *IEEE Communications Surveys & Tutorials*, *15*(1), 387-401. 2013.